\documentclass[a4paper,11pt]{article}
\usepackage{pos}
\usepackage[printonlyused,withpage]{acronym}
\usepackage{booktabs}
\usepackage{caption}
\usepackage[noabbrev]{cleveref}
\usepackage[l3]{csvsimple}
\usepackage{graphicx}
\usepackage[italic]{hepnames}
\usepackage{physics}
\usepackage{siunitx}
\usepackage{subcaption}
\usepackage{todonotes}

\AtBeginDocument{\RenewCommandCopy\qty\SI}

\newcommand{\iu}{\text{i}}

\title{The hadronic decay of vector charmonium}
\ShortTitle{The hadronic decay of vector charmonium}

\author[a]{Beno{\^i}t Blossier}
\author[b]{Jochen Heitger}
\author[a,b]{Jan Neuendorf}
\author*[a,c]{Teseo San José}

\affiliation[a]{
	Laboratoire de Physique des 2 Infinis Irène Joliot-Curie,
	CNRS/IN2P3,\\
	Université Paris-Saclay,
	91405 Orsay Cedex,
	France
}
\affiliation[b]{
	Institut f{\"u}r Theoretische Physik,
	Universit{\"a}t M{\"u}nster,\\
	Wilhelm-Klemm-Str. 9,
	48149 M{\"u}nster,
	Germany
}
\affiliation[c]{
    Higgs Centre for Theoretical Physics, School of Physics and Astronomy,\\
    The University of Edinburgh, Edinburgh EH9 3FD, United Kingdom
}

\emailAdd{benoit.blossier@ijclab.in2p3.fr}
\emailAdd{heitger@uni-muenster.de}
\emailAdd{j\_neue03@uni-muenster.de}
\emailAdd{san-jose-perez@ijclab.in2p3.fr}

\abstract{The extraction of decay parameters using lattice techniques is a computationally expensive task, requiring several volumes and group irreps to relate the spectrum on a lattice simulation to the infinite volume scattering. In this project we employ an alternative method based on a narrow-width approximation to extract the hadronic mixing $\braket*{\APD\PD}{\Pgya}$, which is needed to compute the decay $\Gamma(\Pgya\to\APD\PD)$ between the second excited state of vector charmonium and a pair of $\PD$-mesons in a $p$-wave. We carry out our lattice simulations on two \acs{cls} ensembles at $m_\pi \sim \qty{440}{\mega\eV}$ and $a\sim\qty{0.066}{\femto\meter}$ and obtain results compatible with experiment. Furthermore, we interpret our results analytically using the ${}^3P_0$ quark model.
\newline
\flushright MS-TP-24-37}

\FullConference{%
  The 41st International Symposium on Lattice Field Theory (LATTICE2024)\\
  28 July - 3 August 2024\\
  University of Liverpool, Liverpool, United Kingdom
}

\begin{document}
\maketitle

\section{Introduction}

In these proceedings, we study the hadronic decay $\HepProcess{\Pgya\to\APD\PD}$ of the second excited state of vector charmonium $\Pgya$ into a pair of $\APD\PD$ mesons in a $p$-wave configuration. This process is the main decay channel of $\Pgya$ and takes place close to threshold \cite{ParticleDataGroup:2024cfk}. The standard method to predict a decay width on the lattice relates the spectrum in several finite volumes to the infinite-volume scattering phase-shifts \cite{Luscher:1985dn,Luscher:1986pf,Luscher:1990ux,Luscher:1991cf}. For example, the authors of \cite{Lang:2015sba,Piemonte:2019cbi} have employed this procedure to predict the decay width of several charmonium states, including $\Pgya$.
With our exploratory study carried out in \cite{Blossier:2024dhm} and summarized here, we follow an alternative method developed by Michael and Pennanen \cite{Pennanen:2000yk} that employs ratios of correlators to estimate the decay width. The method, explained in \cref{sec:method}, relies implicitly on a narrow-width approximation, which is satisfied for our decay of interest. Alternative approaches to study hadronic decays include the $\mathcal{K}$-matrix approach \cite{https://doi.org/10.1002/andp.19955070504}, which assumes unitarity and analyticity of the underlying scattering amplitudes, but it also requires $\Ppositron\Pelectron\to\text{hadrons}$ data from experiment in the energy region around the $\Pgya$ pole to fix its free parameters. Several works compute the decay width of $\Pgya$ using this method, see \cite{Hanhart:2023fud} for instance. In our project, we compare our results with the ${}^3P_0$ quark model \cite{MICU1969521,LeYaouanc:1972vsx}. This is a quark rearrangement model which assumes the initial charm quarks are spectators, preserving their quantum numbers and momentum, and a pair of light quarks is created from the vacuum. Then, the four quarks rearrange themselves to obtain the correct quantum numbers of the final state.

Our work predicts the hadronic mixing, $x_{31} \equiv \braket*{\APD\PD}{\Pgya}$, employing the method described in \cite{Pennanen:2000yk}. We establish the connection to the decay width and the energy shift of the states employing non-relativistic quantum mechanics. In particular, we consider the \acs{qcd} Hamiltonian $H_0=\ket*{\Pgya}\bra*{\Pgya}+\ket*{\APD\PD}\bra*{\APD\PD}$ and the perturbation $H_1=\ket*{\APD\PD}\bra*{\Pgya}+\ket*{\Pgya}\bra*{\APD\PD}$. On the lattice, we have a two-level system in the \ac{cm} frame of the charmonium state and the energy shift of the states is obtained solving the eigenvalue problem
\begin{equation}
    \begin{pmatrix}
        -\Delta     & x_{31}   \\
        x^*_{31}    & \Delta
    \end{pmatrix},
\end{equation}
where $\Delta=(m_\psi-E_{\APD\PD})/2$. Including the first-order perturbation, and assuming the charmonium state is heavier, the energies are shifted to
\begin{equation}
    \label{eq:energy-shift}
    m_\psi \to m_\psi + \sqrt{\abs{x_{31}}^2+\Delta^2},
    \qquad
    E_{\APD\PD} \to E_{\APD\PD} - \sqrt{\abs{x_{31}}^2+\Delta^2}.
\end{equation}
Note that in the case of degenerate states with $H_0$, the energy shift is given by the hadronic mixing alone.
The situation in experiment is quite different. Indeed, the decay channel is closed in lattice simulations, as there cannot be a decay in a finite box, and even kinematics may not allow it depending on the quark masses. This means that we can study the effect of the mixing explicitly.
This allows us to test the effects of different theoretical methods. Meanwhile, the situation in experiment corresponds to the decay of a discrete state to a continuum of final states, giving a Breit-Wigner distribution in momentum space and an energy shift in the same direction to both initial and final states.
In particular, we will directly use the hadronic mixing extracted on the lattice to obtain the decay width using Fermi's golden rule. Theoretically, the situation in experiment for non-relativistic states can be understood using \cite{Weisskopf1930}.

\section{Methodology} \label{sec:method}

To extract the hadronic mixing $x_{31} \equiv \braket*{\APD\PD}{\Pgya}$ in finite volume, we consider several vector charmonium interpolators $O_i^\psi$ with $i=1,2,\dots$ Solving a \ac{gevp} yields the real right-eigenvectors $\overline{v}_\alpha$ for the spectrum level $\alpha$,
\begin{equation}
    \label{eq:correlators}
    \begin{aligned}
        & P^{\APD\PD} = \expval{O^{\APD\PD}(t) O^{\APD\PD}(0)},
        &
        & \bar{P}_{\alpha\alpha}^\psi = \bar{v}_{\alpha j} \expval{O_j^\psi(t) O_i^\psi(0)} \bar{v}_{\alpha i},
        &
        & \bar{T}_\alpha = \bar{v}_{\alpha i} \expval{O^\psi_i(t) O^{\APD\PD}(0)}.
    \end{aligned}
\end{equation}
We choose $\alpha=3$ to isolate $\Pgya$, and we assume that the excited vector charmonium and the $\APD\PD$ pair may not be necessarily degenerate, $m_\psi \neq E_{\APD\PD}$. Then, we can study the spectral decomposition of the correlations in \cref{eq:correlators} to isolate $x_{31}$ in terms of the energy difference $\Delta$ \cite{McNeile:2002fh,Blossier:2024dhm},
\begin{equation}
    \label{eq:off-shell-ratio}
    \mathcal{R}(t) = \frac{ \abs{\bar{T}_3(t)} }{ \sqrt{ \bar{P}^\psi_{33}(t) P^{\PD\PD}(t) } }
    \underset{t\gg1}{=} \frac{ \abs{x_{31}} }{ \Delta } \sinh(t\Delta) + Ae^{-t\Delta}.
\end{equation}
If the tuning of the masses is perfect such that $\Delta\to0$, $\mathcal{R}(t) = x_{31}t+A$, where $A$ is a constant. \Cref{eq:off-shell-ratio} is the main relation to extract the mixing, but other relations exist if we transit between ground states with compatible quantum numbers. We can also form the ratio \cite{McNeile:2002fh,Blossier:2024dhm}
\begin{equation}
    \label{eq:xt}
    x_{\text{T}}(t) \equiv
    \dfrac{\bar{T}_3(t)}{\sqrt{\bar{P}^\psi_{33}(t) P^{\PD\PD}}(t)}
    \dfrac{\lambda^{t/2}}{1+\lambda+\dots+\lambda^t} \underset{t \gg 1}{\to} x_{31},
\end{equation}
where $\lambda=\exp(2\Delta)$, and which should tend to a constant if the $\PD$-meson ground state is well isolated without a \ac{gevp}. This is not required for \cref{eq:off-shell-ratio}, and our data do not form clear plateaux to fit \cref{eq:xt}. That is why we prefer to use \cref{eq:off-shell-ratio} for our final results, but also show \cref{eq:xt} in \cref{fig:mixing-vs-p}. Knowing the hadronic mixing, we can compute the decay width assuming that the final $\APD\PD$ are non-interacting, such that each of them can be expressed as a plane wave with \acp{pbc} in a box of size $L$ \cite{McNeile:2002fh,Blossier:2024dhm},
\begin{equation}
    \label{eq:decay-width-lattice}
	\Gamma = \frac{L^3}{24\pi} p_i m_\psi \abs{x_{31}(E=m_\psi)}^2
	\quad \text{with} \quad p_i^2 = \frac{m_\psi^2}{4} - m_{\PD}^2.
\end{equation}
In order for \cref{eq:decay-width-lattice} to tend to the physical result when $L\to\infty$, the hadronic mixing should compensate the factor $L^3$. Note that at no point we establish an explicit connection to the infinite volume decay, and it is clear the method relies on a narrow-width approximation to work. In order to compare our results for $x_{31}$ with the ${}^3P_0$ quark model \cite{LeYaouanc:1972vsx,tayduganov:tel-00648217}, which uses a different normalization for the hadronic mixing, we will compare directly the decay widths. And since both lattice and quark model predict the mixing for any value of the $\PD$-meson momentum, including off-shell kinematics, we lift the condition for energy conservation from the definition of the decay width. In this way we can compare the two theoretical predictions in the entire kinematics, see \cref{fig:decay-width}.

\section{Setup} \label{sec:setup}

\begin{figure}
    \centering
    \begin{subfigure}[c]{0.49\textwidth}
        \centering
        \includegraphics[scale=1]{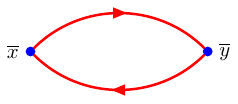}
    \end{subfigure}
    \hfill
    \begin{subfigure}[c]{0.49\textwidth}
        \centering
        \includegraphics[scale=1]{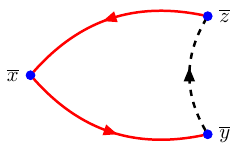}
    \end{subfigure}
    \begin{subfigure}[b]{0.49\textwidth}
        \centering
        \includegraphics[scale=1]{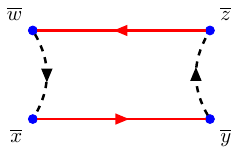}    
    \end{subfigure}
    \hfill
    \begin{subfigure}[b]{0.49\textwidth}
        \centering
        \includegraphics[scale=1]{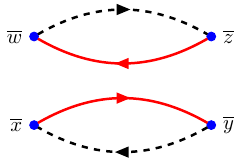}
    \end{subfigure}
    \caption{The Wick contractions considered in this project. The charm quark appears as a solid line and the light-quarks as a dashed line. We do not consider diagrams with charm-quark annihilation.}
    \label{fig:wick-contractions}
\end{figure}

We compute the necessary correlators for \cref{eq:off-shell-ratio} in two \acs{cls} ensembles \cite{Fritzsch:2012wq,Heitger:2013oaa} that employ the Wilson-plaquette gauge action and $N_f=2$ mass-degenerate flavors of non-perturbatively $\order{a}$-improved Wilson quarks. \Cref{tab:cls-nf2-ensembles} gathers the label of the ensembles, their bare-coupling and lattice spacing, the lattice extension, the light-quark hopping parameter, the pion mass, and the value of $m_{\Ppi}L$. A value of $m_{\Ppi}L>4$ is usually required to reduce \acp{fve} to a manageable level. 
\Cref{fig:wick-contractions} shows the Wick contractions considered in this work. 
The red lines indicate a charm-quark propagator, while the dashed lines indicate a light quark. We consider the two charm quarks to belong to two different flavors such that Wick contractions with charm-quark annihilation vanish. To set the momentum of the final-state $\PD$-mesons, we use \acp{ptbc} \cite{Sachrajda:2004mi} on the charm quarks, such that $\Pcharm(n)\to\exp(-\iu a\pmb{n}\cdot\pmb{\theta}/L)~\Pcharm(n)$ and $\APcharm(n)\to\exp(\iu a\pmb{n}\cdot\pmb{\theta}~/L)\APcharm(n)$.
This setup leaves the charmonium state at rest while it boosts the $D$-mesons in opposite directions. Furthermore, leaving the light-quarks with \acp{apbc} preserves isospin symmetry and the final states $\APDzero\PDzero$ and $\PDminus\PDplus$ remain mass-degenerate in our simulations.
We apply the same twist angle $\theta$ in every spatial direction with Gaussian smearing on the quark fields to better interpolate the physical states.
We employ $s=\numlist{20;50}$ smearing iterations for the charm bilinears $\APcharm\gamma_i\Pcharm$, $\APcharm\gamma_4\gamma_i\Pcharm$, $\APcharm\lvec{\nabla}_i\gamma_i\vec{\nabla}_i\Pcharm$, and $\APcharm\lvec{\nabla}_i\gamma_4\gamma_i\vec{\nabla}_i\Pcharm$.
All are simulated at zero momentum and located at the sink.
We used $s=50$ smearing iterations for the $\PD$ meson interpolator $\APcharm\gamma_5\Pup$.
The $\APD\PD$ system at the source is in a $p$-wave,
\begin{equation}
    2O^{\APD\PD} = \APDzero(\pmb{p},t)\PDzero(-\pmb{p},t)-\APDzero(-\pmb{p},t)\PDzero(\pmb{p},t)+\PDminus(\pmb{p},t)\PDplus(-\pmb{p},t)-\PDminus(-\pmb{p},t)\PDplus(\pmb{p},t).
\end{equation}
\begin{table}
    \centering
    \begin{tabular}{cccc ccc}
        \toprule
        {id} & $\beta$ & $a~[\unit{\femto\meter}]$ & $L/a$ & $\kappa_{\ell}$ & $m_{\Ppi}~[\unit{\mega\eV}]$ & $m_{\Ppi}L$ \\
        \midrule
        D5 & 5.3 & 0.0658(7)(7) & 24 & 0.13625 & 449 & 3.6 \\
        E5 & 5.3 & 0.0658(7)(7) & 32 & 0.13625 & 437 & 4.7 \\
        \bottomrule
    \end{tabular}
    \caption[\acs{cls} ensembles]{The two \acs{cls} ensembles used in this work. From left to right, we indicate the ensemble label, the bare coupling constant, the lattice spacing, the number of nodes in every spatial direction ($T=2L$), the light-quark hopping parameter, the corresponding pion mass, and the value of $m_\pi L$.}
    \label{tab:cls-nf2-ensembles}
\end{table}
However, not all Wick contractions can be computed in this way.
The two lower diagrams in \cref{fig:wick-contractions}, which we name box and direct contributions to the $\APD\PD$ propagator, include terms where a momentum change is needed, which is incompatible with \acp{ptbc}.
In this case, we subdivide the correlator into two pieces, one that can be computed using \ac{ptbc}, and another containing the momentum exchange that is computed with Fourier modes and interpolated linearly to the exact momentum.
At $\pmb{p}=\pmb{0}$ both components cancel exactly, while at large momentum the second piece becomes negligible.

\section{Analysis} \label{sec:analysis}

\begin{figure}
    \centering
    \begin{subfigure}[t]{0.49\textwidth}
        \centering
        \includegraphics[scale=1]{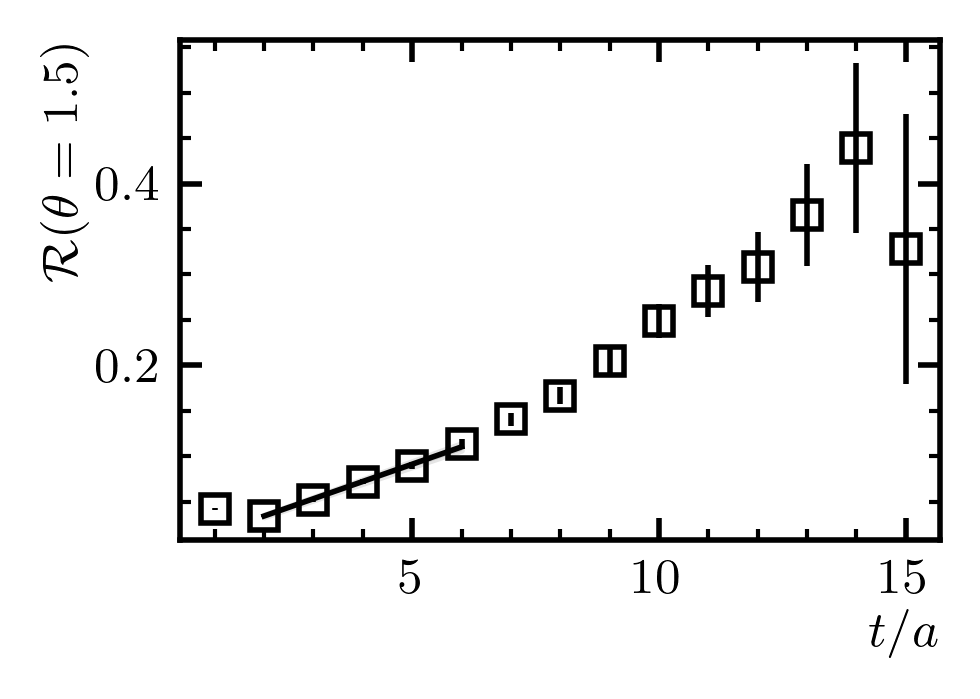}
    \end{subfigure}
    \hfill
    \begin{subfigure}[t]{0.49\textwidth}
        \centering
        \includegraphics[scale=1]{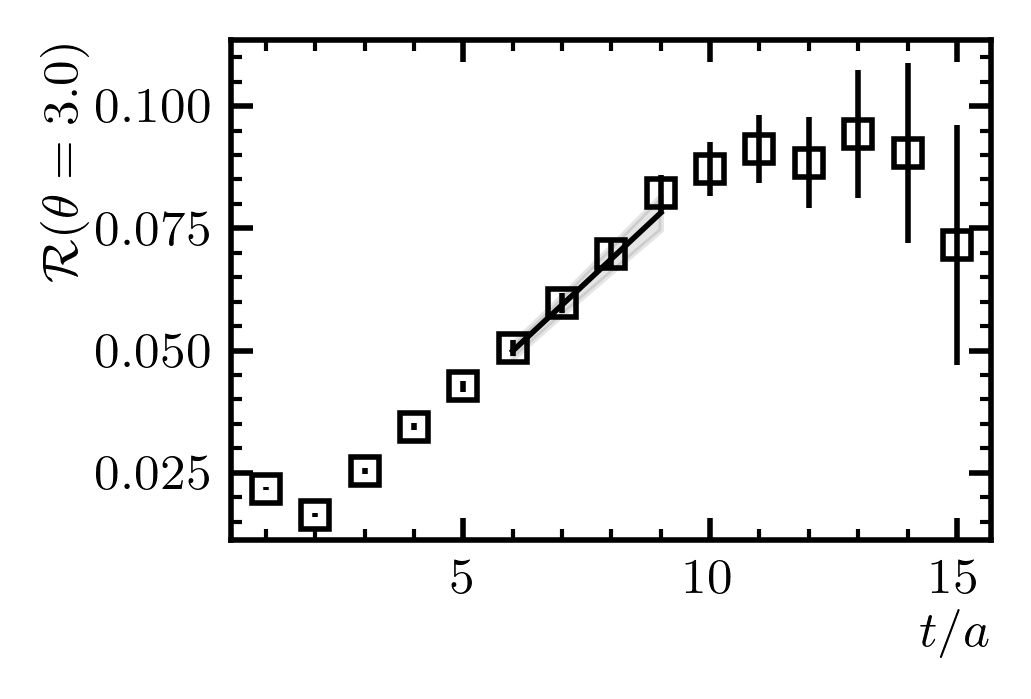}
    \end{subfigure}
    \caption{Ratio $\mathcal{R}(t)$ of \cref{eq:off-shell-ratio} as a function of time, together with fit to the \ac{rhs} of \cref{eq:off-shell-ratio}. Ensemble D5 at twist angle $\theta=\qty{1.5}{\radian}$ (LEFT), and ensemble E5 at $\theta=\qty{3}{\radian}$ (RIGHT). These twist angles lie close to the on-shell condition on their respective datasets, see \cref{tab:results}.}
    \label{fig:fit-slope}
\end{figure}
\begin{figure}
    \centering
    \begin{subfigure}[t]{0.49\textwidth}
        \centering
        \includegraphics[scale=1]{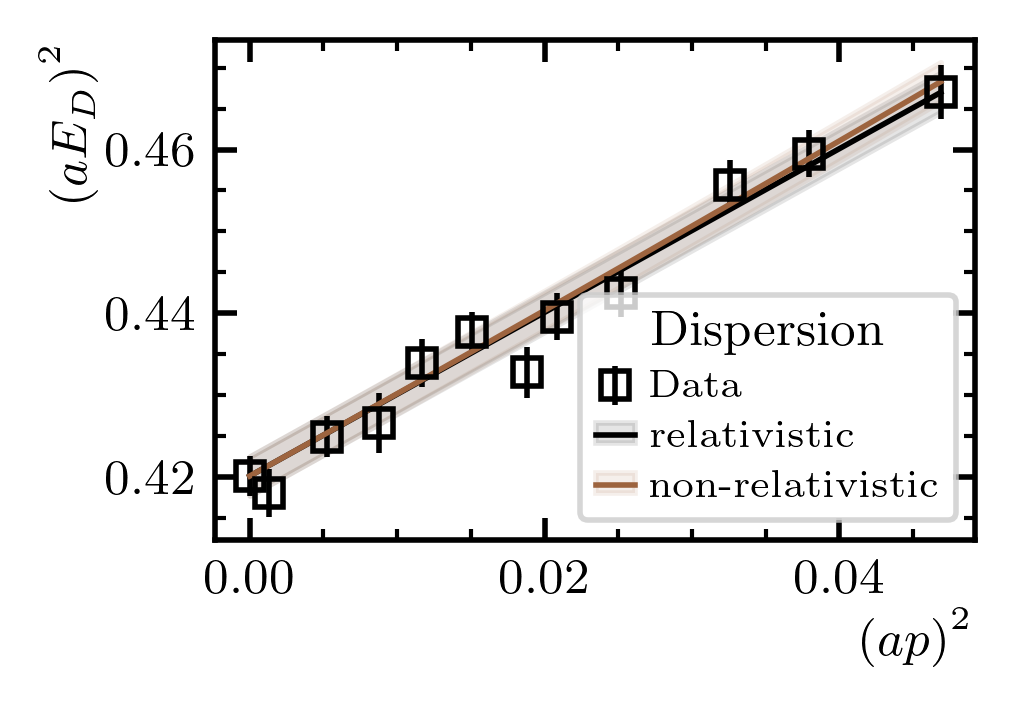}
    \end{subfigure}
    \hfill
    \begin{subfigure}[t]{0.49\textwidth}
        \centering
        \includegraphics[scale=1]{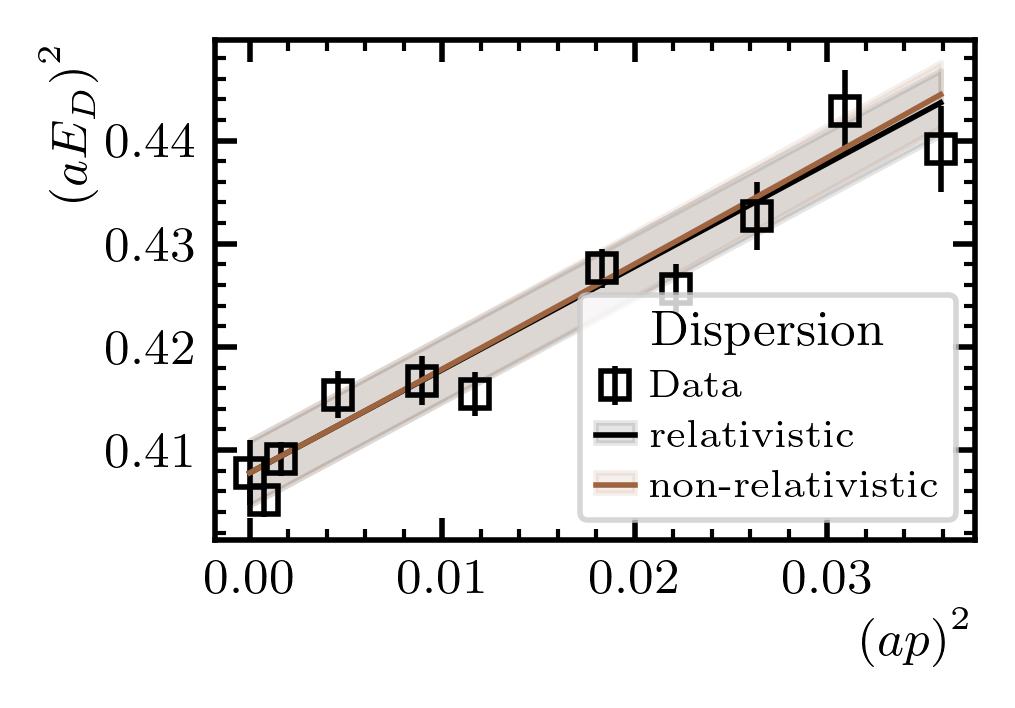}
    \end{subfigure}
    \caption{Dispersion relation for the $\PD$-meson on ensembles D5 (LEFT) and E5 (RIGHT) as a function of the $\PD$-meson momentum modulus, $p\equiv\abs{\pmb{p}}$. The lattice data is fully compatible with a non-relativistic particle in the continuum.}
    \label{fig:d-meson-dispersion}
\end{figure}
\begin{figure}
    \centering
    \begin{subfigure}[t]{0.49\textwidth}
        \centering
        \includegraphics[scale=1]{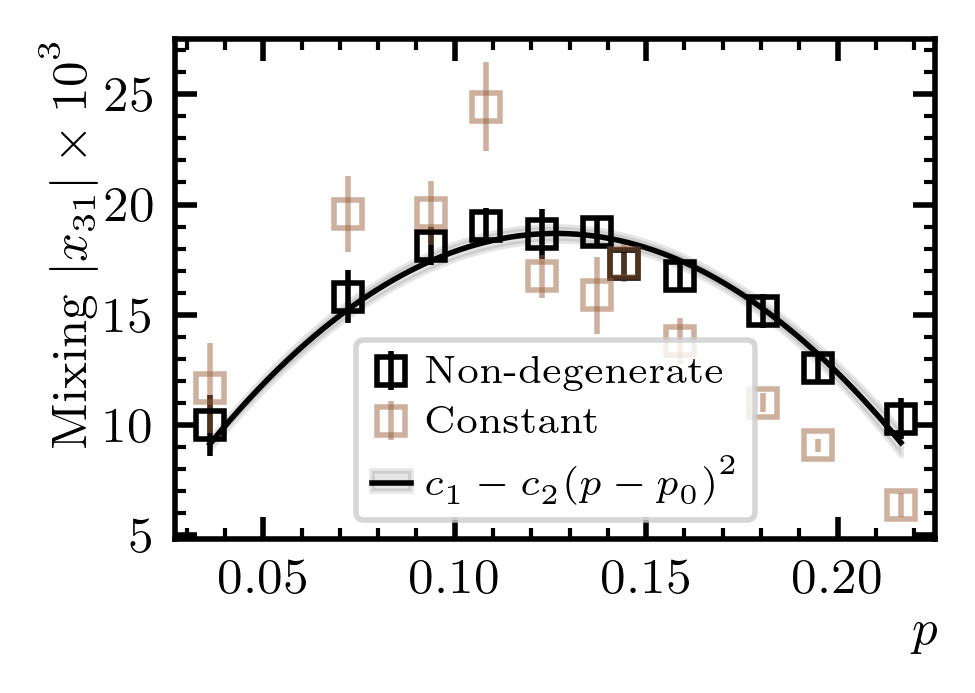}
    \end{subfigure}
    \hfill
    \begin{subfigure}[t]{0.49\textwidth}
        \centering
        \includegraphics[scale=1]{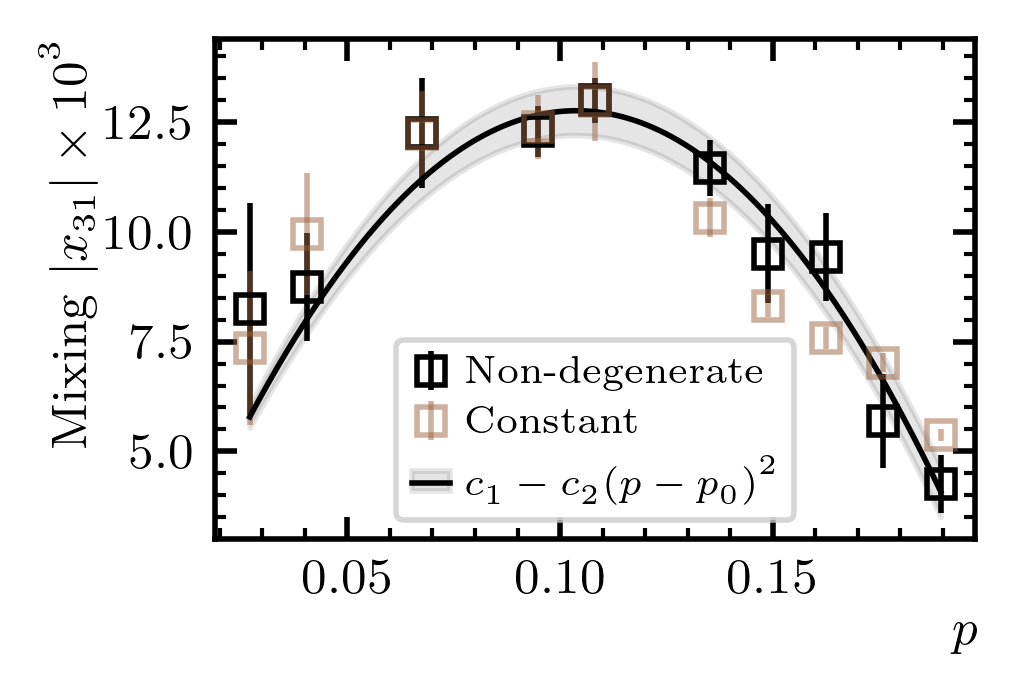}
    \end{subfigure}
    \caption{The hadronic mixing between $\psi(3770)$ and $\APD\PD$ on ensembles D5 (LEFT) and E5 (RIGHT) as a function of the 3-momentum modulus of each $D$-meson. The darker (brighter) points are obtained fitting $\mathcal{R}(t)$ ($x_T(t)$) to \cref{eq:off-shell-ratio} (\cref{eq:xt}). We parametrize the darker points with the parabola in \cref{eq:poly-model}.}
    \label{fig:mixing-vs-p}
\end{figure}
\begin{figure}
    \centering
    \begin{subfigure}[t]{0.49\textwidth}
        \centering
        \includegraphics[scale=1]{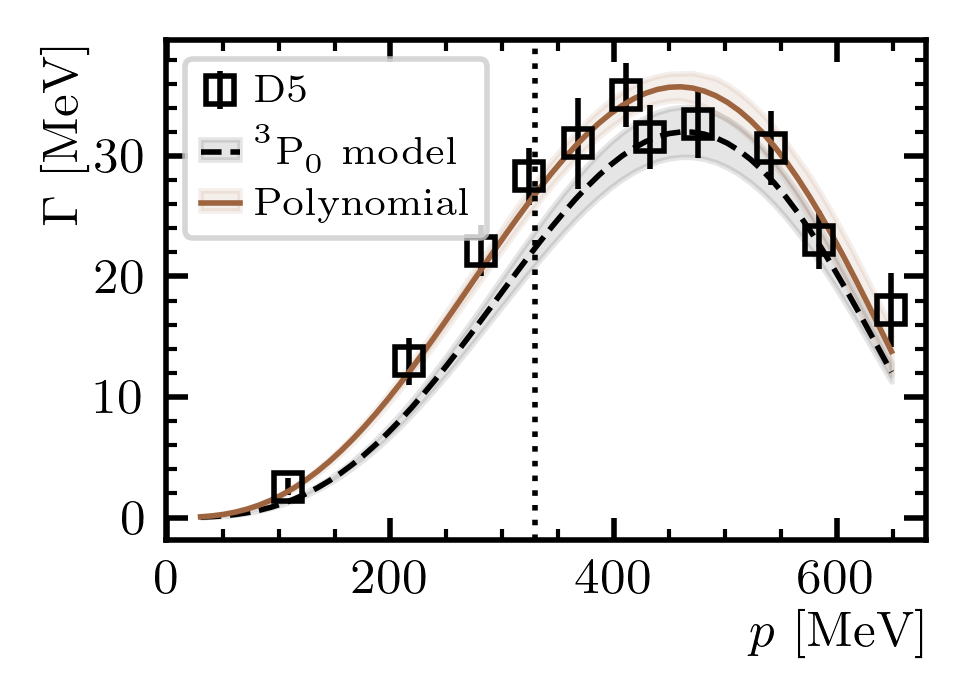}
    \end{subfigure}
    \hfill
    \begin{subfigure}[t]{0.49\textwidth}
        \centering
        \includegraphics[scale=1]{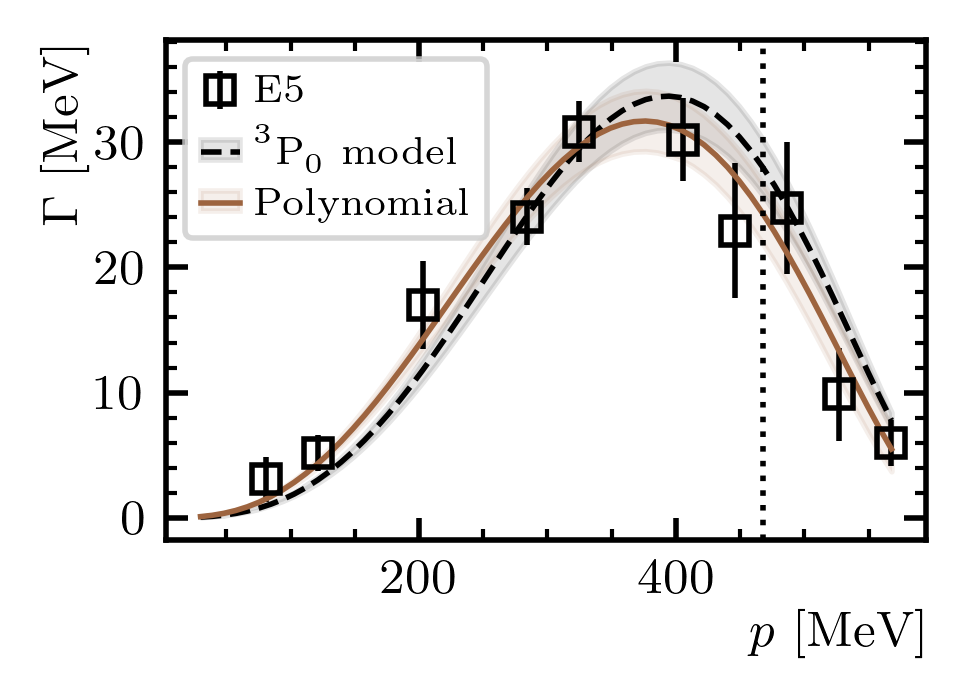}
    \end{subfigure}
    \caption{The decay width $\Gamma(\psi(3770\to\APD\PD)$ as a function of the 3-momentum modulus of each $D$-meson. The slashed line shows the ${}^3P_0$ quark-model prediction, and the full line the prediction given by our lattice determination. The decay may only occur on-shell, marked by the vertical line. We generalize the expression of the decay width for off-shell transitions to compare our results to the quark-model prediction, which employs a different normalization for the hadronic mixing $x_{31}.$}
    \label{fig:decay-width}
\end{figure}

In this section, we extract the mixing $x_{31}$ and compute both the decay width of $\Pgya$ and the energy shift. First, we need to determine the mass of $\Pgya$ in our simulations, $m_\Pgya$. To do that, we solve an $8\times8$ \ac{gevp} with the interpolators and smearings of \cref{sec:setup}. The spectrum is extracted computing the effective mass of the eigenvalues and fitting them to a constant. The fit intervals, quality, and the extracted masses appear in \cref{tab:charmonium-spectrum}. Second, we require the $D$-meson dispersion relation. For each meson momentum, we fit the correlator to a single exponential. In \cref{fig:d-meson-dispersion}, we show the results compared to the free dispersion relation expected in the continuum for both ensembles. These plots indicate the system can be safely treated as non-relativistic. We compute the twist angle corresponding to the on-shell decay assuming the two $\PD$-mesons are noninteracting, so that $m_\Pgya = 2E_{\PD}$ at
\begin{equation}
    \sqrt{3}a\theta_0/L = \sqrt{(am_\psi/2)^2-(am_{\PD})^2}.
\end{equation}
The value in radians for each ensemble appears in \cref{tab:results}. Finally, we can extract the hadronic mixing $x_{31}$ using \cref{eq:off-shell-ratio} or \cref{eq:xt}. \Cref{fig:fit-slope} shows examples of the fits to \cref{eq:off-shell-ratio} on ensembles D5 and E5. All results appear in \cref{fig:mixing-vs-p} as a function of the final $\PD$-meson momentum modulus, $p\equiv\abs{\pmb{p}}$. As mentioned in \cref{sec:method}, \cref{eq:off-shell-ratio} is much better fulfilled in our setup and we employ its output for the final results. Indeed, we observe these points can be fitted to a parabola that vanishes at $p=0$,
\begin{equation}
	\label{eq:poly-model}
	\abs{x_{31}}(p) = c_1 - c_2 (p-p_0)^2,
\end{equation}
where $p = \sqrt{3}\theta/L$, and $p_0=\sqrt{c_1/c_2}$. The results from this fit appear in \cref{tab:hadronic-mixing}. The results employing \cref{eq:xt} agree qualitatively, with a better determination on the larger volume. Plugging \cref{eq:poly-model} in \cref{eq:decay-width-lattice}, we obtain the decay width for the entire kinematical range, see \cref{fig:decay-width}. The on-shell condition is indicated by the vertical dotted line. The main results for an on-shell transition appear in \cref{tab:results}. In particular, we indicate the decay width in lattice and physical units, and the energy shift computed using \cref{eq:energy-shift}. Since ensemble E5 has a larger volume, we choose its values as the main results of this project. In particular, the decay width is compatible with the experiment. Finally, we plot the prediction of the ${}^3P_0$ quark model \cite{tayduganov:tel-00648217} as a grey band in \cref{fig:decay-width}.
We used the results of our spectroscopy calculation to set the parameters of the $\Pgya$ and $\PD$ wave functions.
The remaining free parameter $\beta$, which gives the quark-pair creation strength, is fixed fitting the model to the lattice data, see \cite{Blossier:2024dhm} for more details.
We observe that the quark model describes our data on a qualitative level, and the agreement is especially good on ensemble E5.
This suggests that the physical picture given by the quark model is qualitatively correct.
\begin{table}
	\centering
	\begin{tabular}{l l l l l}
		\toprule
		Level & Fit interval & $\chi^2/\text{dof}$ & $am_{\Pcharm\APcharm}$ & $m_{\Pcharm\APcharm}~[\unit{\giga\eV}]$ \\
		\midrule
		\begin{filecontents*}{tables/D5-charmonium-spectrum.csv}
,Level,FitIntervalA,FitIntervalB,Chisquare,Dof,ChisquareByDof,Pvalue,EnergyLattice,EnergyMeV
1,1,5,20,25.96,15,1.73,0.04,1.02423(63),3074.1(2.4)
2,2,3,12,5.93,9,0.66,0.75,1.286(17),3859(52)
3,3,5,9,0.34,4,0.08,0.99,1.3148(43),3946(13)
4,4,4,8,3.88,4,0.97,0.42,1.379(22),4139(67)
\end{filecontents*}
\csvreader[head to column names, late after line=\\]{tables/D5-charmonium-spectrum.csv}{}
		{\Level & [\FitIntervalA, \FitIntervalB] & \Chisquare/\Dof & \EnergyLattice & $\EnergyMeV$}
		\bottomrule
    \end{tabular}
	\caption{Charmonium spectrum obtained from an $8\times8$ \ac{gevp} on ensemble D5. The effective mass of each eigenvalue is fitted in the interval specified, obtaining the corresponding fit quality. The energy corresponds to the state at rest, given in lattice units and $\unit{\mega\eV}$. We employ the same numbers on ensemble E5.}
	\label{tab:charmonium-spectrum}
\end{table}
\begin{table}
	\centering
	\begin{tabular}{lll lll l}
		\toprule
		Ensemble & $\chi^2/\text{dof}$ & $ac_1$ & $c_2/a$ \\
		\midrule
		\begin{filecontents*}{tables/D5-mixing_vs_theta.csv}
,ChiSquare,Dof,TotalChiSquare,TotalDof,TotalChiSquareByDof,CZero,COne
0,7.16,9,34.39,39,0.88,0.01869(32),1.169(44)
\end{filecontents*}
\csvreader[head to column names, late after line=\\]{tables/D5-mixing_vs_theta.csv}{}{D5 & \TotalChiSquare/\TotalDof & \CZero & \COne}%
		\begin{filecontents*}{tables/E5-mixing_vs_theta.csv}
,ChiSquare,Dof,TotalChiSquare,TotalDof,TotalChiSquareByDof,CZero,COne
0,8.28,8,42.51,35,1.21,0.01276(56),1.183(52)
\end{filecontents*}
\csvreader[head to column names, late after line=\\]{tables/E5-mixing_vs_theta.csv}{}{E5 & \TotalChiSquare/\TotalDof & \CZero & \COne}
		\bottomrule
	\end{tabular}
	\caption{Fit of \cref{eq:poly-model} to the ratio in \cref{eq:off-shell-ratio}.}
	\label{tab:hadronic-mixing}
\end{table}
\begin{table}
	\centering
	\begin{tabular}{lll lll l}
		\toprule
		Ensemble & $\theta_0~[\unit{\radian}]$ & $p~[\unit{\mega\eV}]$ & $a\abs{x_{31}}$ & $\epsilon~[\unit{\mega\eV}]$ & $\Gamma/a$ & $\Gamma~[\unit{\mega\eV}]$ \\
		\midrule
		\begin{filecontents*}{tables/D5-on-shell_decay.csv}
,Ensemble,Momentum,MomentumMeV,Theta,Mixing,Shift,ShiftMeV,DecayWidth,DecayWidthMeV
0,D5,0.110(17),329(52),1.52(24),0.01837(73),0.01837(73),55.1(2.2),0.0089(21),26.8(6.2)
\end{filecontents*}
\csvreader[head to column names, late after line=\\]{tables/D5-on-shell_decay.csv}{}{D5 & \Theta & \MomentumMeV & \Mixing & \ShiftMeV & \DecayWidth & \DecayWidthMeV}%
		\begin{filecontents*}{tables/E5-on-shell_decay.csv}
,Ensemble,Momentum,MomentumMeV,Theta,Mixing,Shift,ShiftMeV,DecayWidth,DecayWidthMeV
0,E5,0.156(13),468(40),2.89(25),0.0095(16),0.0095(16),28.5(4.9),0.0081(21),24.2(6.4)
\end{filecontents*}
\csvreader[head to column names, late after line=\\]{tables/E5-on-shell_decay.csv}{}{E5 & \Theta & \MomentumMeV & \Mixing & \ShiftMeV & \DecayWidth & \DecayWidthMeV}
		\bottomrule
	\end{tabular}
	\caption{The on-shell twist angle, the corresponding momentum modulus of each $D$-meson, the hadronic mixing and mass shift, and the decay width in lattice and physical units.}
	\label{tab:results}
\end{table}

\section{Conclusions and Outlook}

In these proceedings (which summarize our work presented in \cite{Blossier:2024dhm}), we employ the ratio method described in \cref{sec:method} to compute the decay width $\Gamma(\Pgya\to\APD\PD)$ as well as the energy shift in the spectrum. Our main results are gathered in \cref{tab:results}, and in particular we extract a decay width that is compatible with the experimental determination. This project shows that it is possible to extract the decay parameters of an excited state when the narrow-width approximation holds. Our lattice implementation relies on the extraction of the charmonium spectrum via a \ac{gevp} and the use of \ac{ptbc} to tune the $\APD\PD$ system in a $p$-wave. Our determination of the hadronic mixing $\braket*{\APD\PD}{\Pgya}$ is not limited to on-shell kinematics, and we compare our results to the prediction by the ${}^3P_0$ quark model, which shows remarkable agreement, all while being an analytical method to deduce the decay parameters. This suggests that the physical picture presented by the quark-rearrangement model is qualitatively correct. Although we limit ourselves to two ensembles with different volumes, the method presented here can be extended to a more complete dataset. In particular, an extrapolation to the continuum limit in the near future is reasonable. In a similar fashion, lowering the light-quark masses to the physical point should have only a limited effect.

\acknowledgments

The work of TSJ is supported by the Agence Nationale de la Recherche under the contract ANR-17-CE31-0019, and by the STFC Consolidated Grant ST/X000494/1 Particle Theory at the Higgs Centre. The work of JN and JH is supported by the program “Netzwerke 2021”, an initiative of the Ministry of Culture and Science of the State of Northrhine Westphalia, in the NRW-FAIR network, funding code NW21-024-A. Additionally, JN and JH were supported by the German Research Foundation (DFG) through the Research Training Group “GRK 2149: Strong and Weak Interactions – from Hadrons to Dark Matter”.
GENCI granted access to the HPC resources of TGCC (2024-A0160502271). Parts of the calculations were performed on the HPC cluster PALMA II of the University of Münster, subsidized by the DFG (INST 211/667-1). JH thanks the long-term workshop on HHIQCD2024 at the Yukawa Institute for Theoretical Physics (YITP-T-24-02) for giving him a chance for useful discussions as this work was finished. The authors thank Alain Le Yaouanc and Antonin Portelli for fruitful discussions.

\acrodef{apbc}[anti-PBC]{anti-periodic boundary condition}
\acrodef{ape}[APE]{Array Processor Experiment}
\acrodef{cg}[CG]{Clebsch-Gordan}
\acrodef{cls}[CLS]{Coordinated Lattice Simulations}
\acrodef{cm}[CM]{centre-of-mass}
\acrodef{chpt}[ChPT]{chiral perturbation theory}
\acrodef{ddhmc}[DD-HMC]{domain decomposition hybrid Monte Carlo}
\acrodef{fve}[FVE]{finite-volume effect}
\acrodef{gevp}[GEVP]{generalized eigenvalue problem}
\acrodef{lqcd}[LQCD]{lattice QCD}
\acrodef{pbc}[PBC]{periodic boundary condition}
\acrodef{ptbc}[PTBC]{partially twisted boundary condition}
\acrodef{qcd}[QCD]{quantum chromo dynamics}
\acrodef{rhs}[RHS]{right-hand side}
\acrodef{tbc}[TBC]{twisted boundary condition}

\bibliographystyle{JHEP}
\bibliography{biblio}

\end{document}